# Dependence of the solar wind plasma density on moderate- and extremely high-geomagnetic activity elucidated by potential learning

Ryozo Kitajima[1*†], Motoharu Nowada[2*†] and Ryotaro Kamimura[3]

**Abstract**

In this study, the relationship between moderate and extremely high levels of geomagnetic activity, represented by the $K_p$ index (2- to 5+and 6- to 9), and solar wind conditions during southward IMF intervals was revealed utilizing a newly developed machine learning technique. Potential learning (PL) is a neural network algorithm that emphasizes input parameters with the highest variance during training and identifies the most significant input parameters influencing the outputs based on a computed metric called "potentiality". We focus on discussing the dependence of solar wind plasma density on moderate-geomagnetic conditions. It has poorly been understood from what stage of geomagnetic activity the solar wind density begins to control the $K_p$ level. Previously, PL extracted the solar wind velocity as the most predominant parameter at extremely low (0 to 1 +)- and high-$K_p$ ranges under southward interplanetary magnetic field (IMF) conditions. Also in this study, the IMF three components, solar wind flow speed, and plasma density obtained from the OMNI solar wind database (1998–2019), corresponding from solar cycle 23 to beginning of cycle 25, were used as the input parameters. Again, PL selected the solar wind velocity as the most significant parameter for the moderate and extremely high $K_p$ levels. The potentiality of solar wind density for these $K_p$ ranges was, however, 3.5 times higher than that in the previous study, suggesting that its impact on geomagnetic activity cannot be ignored. We statistically investigated the relation between solar wind speed and plasma density used as the PL input data under all $K_p$ levels. At higher than the moderate $K_p$ level, geomagnetic conditions become high even under slow solar wind velocity, if the plasma density is large, suggesting that not only solar wind velocity but also plasma density might significantly contribute to geomagnetic activity. These PL and incidental statistical investigations show that the solar wind density begins to regulate $K_p$ higher than moderate geomagnetic activity level under southward IMF conditions. They also would greatly help not only understand general relationship between solar wind conditions and geomagnetic activity but also forecast geomagnetic activity under various IMF conditions.

**Keywords** Space weather modeling, Solar wind conditions, Geomagnetic activity, Machine learning (potential learning)

† Ryozo Kitajima and Motoharu Nowada have equally contributed to the paper.

*Correspondence:
Ryozo Kitajima
r.kitajima@eng.t-kougei.ac.jp
Motoharu Nowada
moto.nowada@sdu.edu.cn
Full list of author information is available at the end of the article

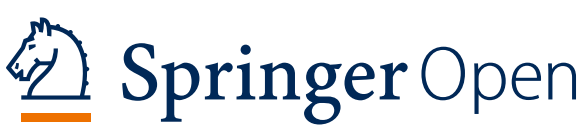

## 1 Introduction

The terrestrial magnetosphere is always exposed to the high-speed particle flows originating from the sun (referred to as solar wind) and changes dynamically because of the interactions with solar wind plasmas and magnetic fields, so-called interplanetary magnetic field (IMF, e.g., Black 1967; Glassmeier et al. 2009; Glassmeier and Vogt 2010). Therefore, geomagnetic disturbances strongly depend on solar wind conditions, in particular, the IMF conditions. Magnetic reconnection occurring on the boundary region between geomagnetic field and IMF is the main driver enhancing the geomagnetic activity because this process is caused by the breaking down of the geomagnetic field frozen-in condition, caused by reconnecting the geomagnetic field with the field lines of IMF (Parker 1957; Sweet 1958; Petcheck, 1964), and leads to magnetic disturbances in the magnetosphere (e.g., Dungey 1961). Given that dayside magnetosphere where the IMF and geomagnetic field first encounter has dominant northward magnetic field, the magnetic reconnection occurrence is high when the IMF orientation is southward.

The $K_p$ index is a widely accepted parameter for measuring the global geomagnetic activity, which can be obtained based on the weighted average of the geomagnetic activity indices ($K$ indices) at 13 geomagnetic observatories worldwide (Bartels 1949) with a 3 h time resolution. It has been well known that there is good correlation between $K_p$ and solar wind parameters at L1 (Lagrange) point, which were revealed by Wing et al. (2005), Wintoft et al. (2017), Zhelavskaya et al. (2019), Shprits et al. (2019), and references therein. Newell et al. (2008) succeeded in formulating the $K_p$ index which has the functions of the IMF and energy coupling functions, which show quantitative entry amount of solar wind electromagnetic energy to the magnetosphere using the IMF and solar wind plasma moments as proposed by Newell et al. (2007). An early study of Snyder et al. (1963) and recent Elliott et al. (2013) statistically revealed that a close relationship exists between $K_p$ and the solar wind velocity. Particularly, Elliott et al. (2013) revealed the detailed relation between $K_p$ and solar wind velocity measured under three difference solar wind conditions, such as solar wind compression, rarefaction, and interplanetary coronal mass ejections (ICMEs) to improve the accuracy of the $K_p$ predictions. They aimed to increase accuracy for the geomagnetic activity predictions by tracking various solar wind structures with the solar, coronal, and heliospheric imaging. Nevertheless, it has been difficult to estimate the geomagnetic activity level using the solar wind parameters.

Approaches using machine learning (or deep learning) technique, such as neural network (NN) with the input parameters of the IMF and solar wind plasma, have been adopted to predict geomagnetic activity (e.g., Boberg et al. 2000; Chakraborty and Morley 2020; Conde et al. 2023; Wing et al. 2005). Recently, in addition to solar wind data, training machine learning models on long-term relationship between solar surface conditions which are derived from solar images and corresponding geomagnetic activity index ($K_p$) has enabled more precise predictions of short-term geomagnetic disturbances (activity) (Bernoux et al. 2022; Wang J et al., 2023; Wang T et al., 2025). To predict the $K_p$ index, based on the three-hour averaged values of the IMF and solar wind plasma parameters, Boberg et al. (2000) developed a multilayer feed-forward network with the hybrid model evaluated in terms of "training", "validation", and "test" with the correlation coefficient (CC) and root-mean-square error (RMSE). However, there were discrepancy between measured and predicted $K_p$ when the solar wind parameters attributed to the occurrence of geomagnetic storm were utilized as NN inputs. They concluded that it was difficult to predict the entire broad $K_p$ range from 0 to 9, but the wide range $K_p$ predictions may be possible, if the two evaluated models in their study can become one hybrid model. Bala and Reiff (2012) exploited the $K_p$ index prediction model based on a NN and compared several $K_p$ predicting patterns with inputs of various IMF and solar wind plasma conditions. In particular, they took into account the energy transfer rate from solar wind which can be represented by solar wind coupling functions, proposed by Boyle et al. (1997) and Newell et al. (2008), as the input parameters. The most significant difference between these two coupling functions is the inclusion of the solar wind density term in their formulae. The solar wind density is believed to play a significant role in effectively transferring the solar wind energy to the magnetosphere (e.g., Lopez et al. 2004; Kataoka et al., 2005), and resultantly, its variation is considered to control the geomagnetic activity level. To accommodate the solar wind density term in the NN inputs, they introduced the Newell's solar wind coupling function which includes the dynamic pressure term, and compared the $K_p$ predictions with the NN inputs between Boyle and Newell coupling functions. The result, however, showed that the dynamic pressure term was not much effective in accuracy for the prediction of $K_p$ that represents global geomagnetic activity level. Therefore, solar wind density might be considered as a direct NN input parameter when predicting the global geomagnetic activity. They also identified significant differences in the RMSE and correlation coefficients between the models.

Another machine learning technique, such as a support vector machine (SVM), was applied to build a $K_p$ prediction model by Ji et al. (2013). The performance of their SVM-based model was evaluated by comparisons with the $K_p$ prediction models which were developed based on a NN. They constructed prediction models for the $K_p$ index exceeding 6. In their study, the NN- and SVM-based 4 machine learning models for the intense $K_p$ (larger than 6) predictions were created and compared the performance between them. The constructed 4 machine learning models are (1) a NN model using only solar wind input data, (2) a SVM model with solar wind data, (3) a NN model using solar wind and preliminary $K_p$ data computed based on the geomagnetic field measured at the specific 8 ground magnetic observatories, and 4) a SVM model with solar wind and preliminary $K_p$ index data. As a result, a NN model with the input data of solar wind and preliminary $K_p$ showed the best accuracy for the prediction of $K_p$ larger than 6. Unfortunately, it seems that a SVM technique did not had a good accuracy for the intense $K_p$ prediction, but it may be effective if the preliminary $K_p$ data cannot be obtained, or the $K_p$ index lower than 6 is required to predict, based on the solar wind data.

Tan et al. (2018) developed and evaluated a prediction model for the low $K_p$ level and intense geomagnetic conditions ($K_p \geq 5$) which are corresponding to non-geomagnetic and geomagnetic storm intervals. Furthermore, their model took into account the prediction error of the $K_p$ index with long short-term memory (LSTM), constructed by recurrent NNs (RNNs; Hochreiter and Schmidhuber 1997). They also used the solar energy input function (solar wind coupling function) and associated viscous term, proposed by Newell et al. (2008), and the historical $K_p$ index, as the input parameters of the solar wind conditions besides the IMF and solar wind plasma data, as well as Bala and Reiff (2012), Ji et al. (2013), and Conde et al. (2023). Investigating the prediction performance of $K_p$ during the intervals with/without geomagnetic storms, based on their LSTM NN models, in terms of the statistical parameters, such as RMSE, CC, and mean absolute error (MAE), the values of the statistical parameters in their model were better than those in the $K_p$ predictions using the other machine learning models, that is, the SVM model and the models proposed by Boberg et al. (2000) and Bala and Reiff (2012). Because the values of error and CC in the LSTM-based $K_p$ predictions during the intense geomagnetic activity were particularly smaller but larger than those in the other models, the LSTM-based model is expected to be a useful $K_p$ prediction model with high accuracy and low errors under geomagnetic disturbance conditions.

Kitajima and Nowada et al. (2022; hereafter referred to as KN2022) defined $K_p$ = 0 to 1 + as extremely low geomagnetic activity level and $K_p$ of 6 + to 9 as the extremely high geomagnetic activity level and examined solar wind conditions classifying them under southward IMF conditions using 22 years' solar wind OMNI and $K_p$ databases. In their study, a newly developed NN, potential learning (PL), was applied for the first time to time series data in space plasma, and a $K_p$ classification model was constructed to extract the most significant solar wind parameter, that is, input parameter with a value of the highest potentiality that would drive large geomagnetic disturbances. They succeeded in extracting the solar wind velocity as the most influential parameter that is closely related to extremely low- and high-$K_p$ conditions. Their results obtained by PL were also consistent with the relation between solar wind parameters and $K_p$, which can be described by the empirical equations derived by statistical data analysis (Newell et al. 2008).

Although this study is an extension of KN2022, we defined $K_p$ = 2- to 5 + as moderate geomagnetic activity level adding to the extremely low- and high- geomagnetic activity and discuss solar wind conditions classifying moderate- and extremely high-geomagnetic activity conditions under southward IMF. Furthermore, by extracting the most significant solar wind parameter(s) with PL and comparing the results obtained in this study with those from KN2022, we elucidated the ultimate solar wind parameter(s) governing the geomagnetic activity under southward IMF conditions that the magnetosphere would be largely disturbed.

The data, compilation, and methodology used in this study are described in Sect. 2. The results obtained based on the PL and incidental statistical analyses using the PL database are presented in Sect. 3. The discussion and summary are provided in Sect. 4.

## 2 Data and methodology

### 2.1 Analysis of large databases

In this study, we attempted to clarify the relationship between solar wind conditions, and the moderate and extremely high $K_p$ levels by developing a model that utilizes IMF and solar wind plasma parameters as inputs and classifies $K_p$ according to its magnitude (level). The analysis method used in this paper is PL, which is main analytical tool utilized in KN2022.

PL was basically created to identify the highest significance among the input parameters (e.g., Kamimura 2017) and has been applied to address the issue in field of marketing research (e.g., Kitajima et al. 2016). In KN2022 and this study, we attempt to find what the solar wind parameters contribute essentially to cause extremely low-, high-, and moderate-geomagnetic conditions under southward IMF, and then PL is so useful to identify and extract them.

PL has an algorithm with the two stages: knowledge accumulation, based on self-organizing map (SOM),

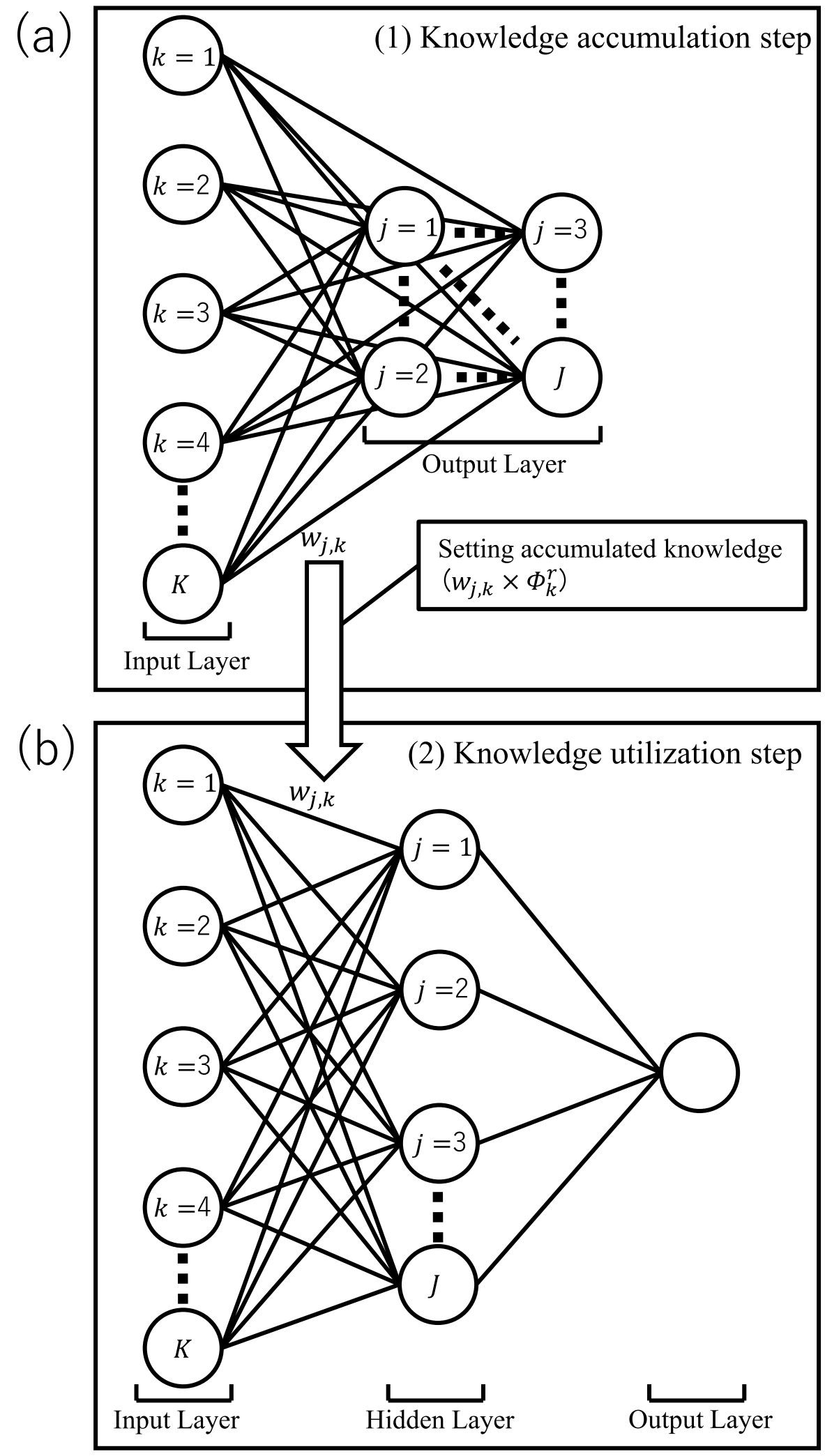

**Fig. 1** The concept and structures of potential learning (PL) are shown. Two important steps (**a**. knowledge accumulation and **b**. knowledge utilization) at PL are illustrated

illustrated in Fig. 1a, and knowledge utilization originating from multilayer perceptron (MLP) as shown in Fig. 1b. SOM is a type of unsupervised neural network, which was utilized as a function to make an initial value for weight values used in knowledge utilization step in PL. At the stage of knowledge accumulation, first, the input neuron's potentiality is calculated, and associated knowledge is acquired (training). We, here, introduce "potentiality" as ability to respond to various conditions of neuron. The higher potentiality of the input neuron is, the neuron becomes more important at the training stage. In PL, we can interpret which input parameters are significant by focusing on their potentiality values after training. Note that the variance value of the input neuron can almost be treated as that of each input parameter.

If the number $k(k = 1, 2, \ldots, K)$ is assigned to the input neuron, we can obtain the potentiality of the $k$th input neuron $\left(\Phi_k^r\right)$ between 0 and 1 by the following equation:

$$\Phi_k^r = \left( \frac{V_k}{\max\limits_{k=1,\ldots,K} V_k} \right)^r \tag{1}$$

, where $V_k$ is the variance of the $k$th input neuron which can be computed, based on weight values $\left(w_{j,k}\right)$ connected to the $k$th input neuron from $j$th $\left(j = 1, 2, \ldots, J\right)$ output neuron. The value of $r$ is a parameter which emphasizes an input with high potentiality. The larger $r$ becomes, the large variance input parameter can be identified as the significant input parameter with large potentiality.

PL was trained with SOM, where the potentiality was used to calculate the distance ($d_j$) between the input neuron (the input from the $k$th input neuron is shown with $x_k$) and the $j$th output neuron using the following formula after the input neuron's potentiality was derived:

$$d_j = \sqrt{\sum_{k=1}^{K} \Phi_k^r \left(x_k - w_{j,k}\right)^2}. \tag{2}$$

Equation (2) shows that the distance weighted by the potentiality of the input neuron was used at the training stage. The training logic was the same as that of the SOM. PL starts the MLP-based training at the step of knowledge utilization. In this stage, the weight obtained in the knowledge accumulation stage was multiplied by the potentiality and was set as the initial weight between the input and hidden layers for learning. The MLP-based training depends on the initial weights. PL is, however, expected to provide more precise training than the MLP-based training, utilizing the knowledge obtained from the input parameters.

In the knowledge utilization step shown in Fig. 1b, hyperbolic tangent function and softmax function as defined with exp(input)/Σ(exp(input)) were adopted to the activation functions of the hidden and output neurons, respectively. To search the most suitable r value in Eqs. (1) and (2), we varied its value from 1 to 10 with a step of 1, and also calculated the associated accuracy (see supplementary information on how to compute accuracy and Sect. 3.1 described later). A total of 1,126 (70%) of the 1,606 samples was used for training. Half of the remaining 240 samples (15%) was utilized to prevent training from overfitting (early stopping) and the other half (15%) was used for testing. These allocation rates which were assigned to training and testing in the PL are the same as those in KN2022. As shown in Fig. 2, we maintained them during the PL runs and made 10 different models in a random choice manner. Furthermore, the performance of each model was evaluated, based on computations of the average values of the 10 models.

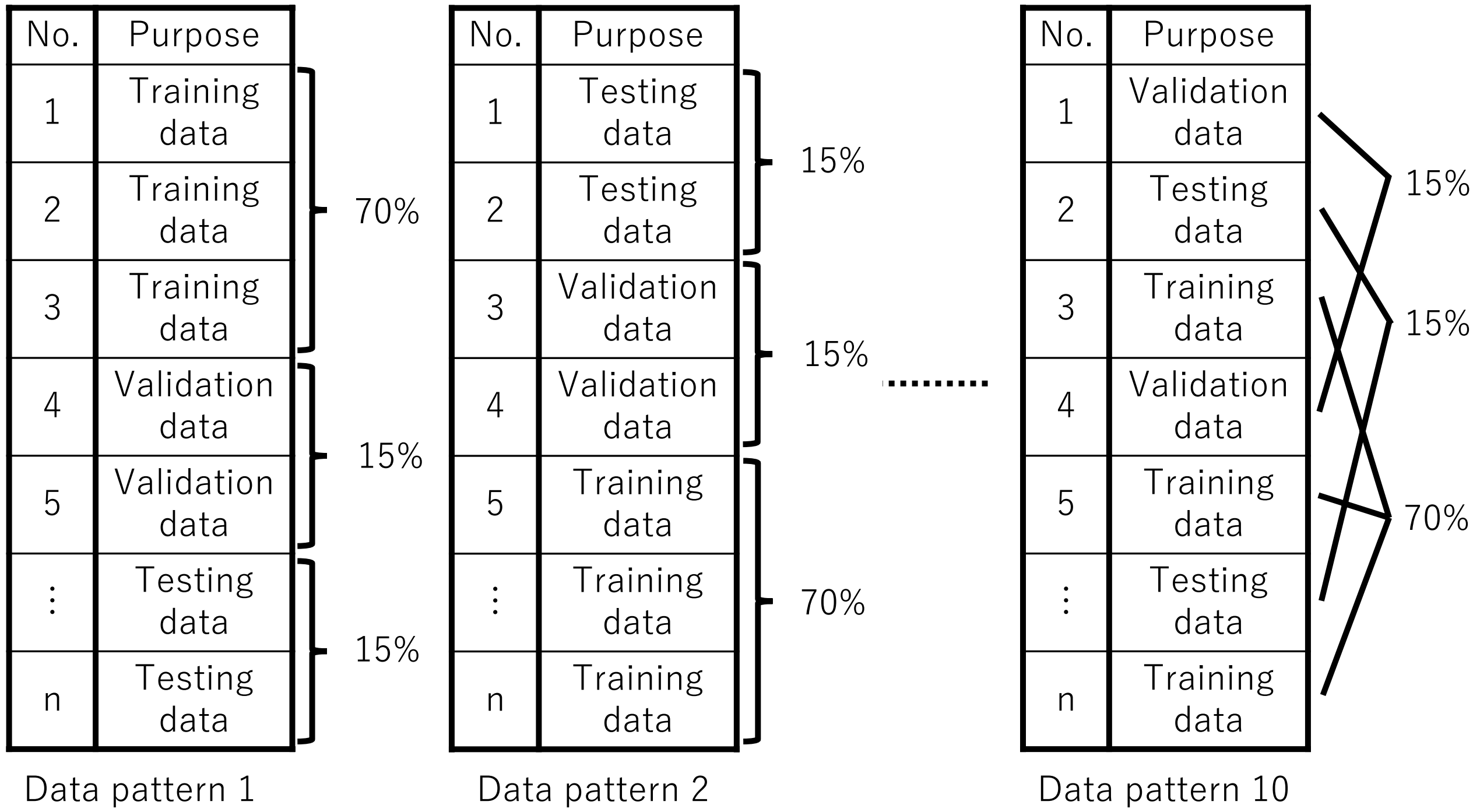

**Fig. 2** Block diagrams show the details of 10 potential learning (PL) models. In each model, the data for training (training data), the data to prevent training from overfitting (early stopping, validation data), and the data for testing the model (testing data) are included. The rates for these three types of data in the PL are 70%, 15%, and 15%, respectively

PL is a machine learning technique to extract the most influential input parameter for the output parameters (parameters with high potentiality). We, however, sometimes need to perform statistical investigations, based on the database used in PL, to physically understand how the high-potentiality parameters significantly affect moderate- and extremely high-geomagnetic activity, and whether or not they also have significant relationship with each other at different levels of $K_p$. Furthermore, note that the PL result does not always directly reflect the result obtained from statistical investigation, because PL is the result based on large/small variance of the parameter, but the statistical results are not.

PL consisted of five input and 192 output neurons in the knowledge acquisition phase and 192 intermediate and 2 output neurons in the prediction phase, which were determined by the sampled data size. The transfer function of intermediate neurons is a hyperbolic tangent function in the prediction phase and that of output neurons is a softmax function.

**Table 1** Input solar wind parameters for PL

| No | Input paramaters | Unit |
|---|---|---|
| 1 | $B_X$ [IMF GSM-X component] | nT |
| 2 | $B_Y$ [IMF GSM-Y component] | nT |
| 3 | $V_X$ [Solar wind velocity in GSM] | km/s |
| 4 | $N_P$ [Ion Number Density] | $/cm^3$ |
| 5 | $B_S^\dagger$ [Southward IMF GSM-Z Component] | nT |
| | $B_S^\dagger = \begin{cases} 0 & (B_Z > 0), \\ B_Z & (B_Z \leq 0). \end{cases}$ | |

Information on input (solar wind) parameters used in this study is given. These input parameters were normalized in the range between 0 and 1 ($x'$) based on the following equation: $x' = (x - \min(x))/(\max(x) - \min(x))$.

## 2.2 Database compilation

We used a large OMNI solar wind and geomagnetic activity index databases including records from 22 years, that is, from January 1, 1998, to December 31, 2019, corresponding from solar cycle 23 to the beginning of cycle 25. The OMNI solar wind data have a one-minute time resolution and are time-shifted from the satellite's solar wind observation location to the bow shock nose. Table 1 summarizes the five parameters to characterize solar wind conditions for input of classifier built based on the PL algorithm. The IMF three components include the sun-earth directional component (IMF-$B_X$), the component of dawn–dusk direction (IMF-$B_Y$), and southward component (IMF-$B_S$), respectively. These parameters were normalized in the range between 0 and 1 ($x'$), which can be derived from the following equation: $(x - \min(x))/(\max(x) - \min(x))$, where x is input parameter, and min(x) and max(x) are minimum and maximum input values, respectively. The time resolutions of the solar wind parameters and the $K_p$ index are 1 min and 3 h, respectively. We adjusted the time resolution difference between solar wind data and $K_p$ by calculating the 3-h average of solar wind data. If the number of solar wind data met less than 40% of the total, no average of the data is taken. We used only in situ solar wind observation data, when the GSM-X component (sun-earthward directional component) of the satellite location was larger than the nominal bow shock nose location (~15 $R_E$), derived from the bow shock model proposed by Farris and Russell (1994). This satellite location threshold must be established to completely exclude IMF and solar wind plasma data from the OMNI database, as solar wind observation satellites occasionally measure within the magnetosphere. Furthermore, we only used solar wind conditions and geomagnetic activity during southward (negative) IMF-$B_Z$ intervals as input parameters for PL, labeled as $B_S$ in Table 1. This is because geomagnetic conditions are favorable to be moderately and highly disturbed with high occurrences of magnetic reconnection at the dayside magnetosphere under southward IMF conditions (e.g., Dungey 1961). Although the intensity of southward IMF component ($B_S$) is a fundamental parameter to cause the disturbances in the magnetosphere, it is important to investigate whether or not the IMF-$B_S$ intensity alone plays a dominant role in causing geomagnetic disturbances and/or determining the disturbance scale. This is a reason why we selected the IMF-$B_S$ as a PL input.

Before the solar wind parameters were inputted to the PL classifier, the solar wind data were categorized into positive and negative groups (targets), based on the $K_p$ levels. The solar wind data associated with the extremely high-$K_p$ (6- to 9) range were categorized as positive target (group). In contrast, the solar wind parameters were classified as negative target (group) when the corresponding $K_p$ index shows moderate geomagnetic activity level (2- to 5+). In KN2022, extremely low $K_p$ indices and the associated solar wind parameters were set in negative target (group) of this PL classifier.

The total data point number was 30,541, including a positive (negative) target number of 803 (29,738). We, however, selected randomly 803 points from 29,738 negative target data points, in order to equalize the data numbers between positive and negative targets. Finally, 1,606 positive and negative data points were utilized for the PL classifier. In this study, we focus on in situ observations for both positive and negative target events, which enables us to extract significant parameters directly from raw, unprocessed databases. Therefore, we did not employ oversampling techniques that duplicate existing positive samples or generate synthetic ones to increase the number of positive samples and address imbalance of positive and negative targets.

## 3 Results

### 3.1 Model evaluation

We calculated the values of four measures (accuracy, precision, recall, and F-measure) and evaluated the $K_p$ classification model using PL. Definitions and equations to compute the values of four measures are summarized and explained in supplementary information (additional) file. Note that as accuracy value particularly approaches 1, the PL model performance is considered to be better. The four measures computed while varying the value of "*r*" from 1 to 10, as well as those obtained from the MLP, are listed in Table 2. Table 2 shows the calculation results of the four measures for extremely low- and high-$K_p$ cases, as shown in KN2022, and those obtained in this study to compare the reliability of this classification model with that of the previously reported one. In the KN2022 case, the value of r in the equation to derive the potentiality [see Eq. (3) in KN2022] was 5 and the accuracy value was 0.9875, which was slightly smaller than that in multilayer perceptron (MLP). We adopt $r=2$ as the parameter value, as it achieved an accuracy of 0.9100, which was the highest among all values of "*r*" ranging from 1 to 10 (Table 2). This accuracy was also smaller than that in MLP, while its value exceeded 0.9000 as well as that in KN2022. The models built using PL demonstrated high accuracy. We chose the PL algorithm rather than MLP because it allows us to identify and extract the most influential input parameter on the output parameter, even though the precision value in PL (0.9276) is only slightly larger than that of MLP (0.9232).

### 3.2 Statistical distributions of solar wind plasma

The occurrence rate histograms of the solar wind velocity ($V_X$), plasma density ($N_P$), IMF-$B_S$ (southward $B_Z$) which were used as the input data of PLs in KN2022 and this study are shown in Figs. 3, 4, and 5, respectively. The bin sizes used in these three figures are 100 km/s, 1.0/cc, and 1.0 nT, respectively. In Fig. 5, to clearly show the occurrence rates of all IMF-$B_S$ ranges within each $K_p$ group, we added magnified histograms of the IMF-$B_S$ occurrence rates.

In Fig. 3, a velocity distribution peak can be found in the range of 300–400 km/s at extremely low geomagnetic

**Table 2** Four measures for "*r*" ranging from 1 to 10, as well as those from the MLP, and comparison of the results between KN2022 and this study

| | PL | | | | | | | | | | MLP |
|---|---|---|---|---|---|---|---|---|---|---|---|
| | ***r*** **= 1** | ***r*** **= 2** | ***r*** **= 3** | ***r*** **= 4** | ***r*** **= 5** | ***r*** **= 6** | ***r*** **= 7** | ***r*** **= 8** | ***r*** **= 9** | ***r*** **= 10** | |
| Accuracy | 0.9096 | **0.9100** | 0.9067 | 0.9017 | 0.9079 | 0.9071 | 0.9083 | 0.9092 | 0.9092 | 0.9079 | **0.9213** |
| Precision | 0.9262 | **0.9276** | 0.9211 | 0.9191 | 0.9230 | 0.9210 | 0.9234 | 0.9274 | 0.9216 | 0.9227 | 0.9232 |
| Recall | 0.8908 | 0.8900 | 0.8900 | 0.8817 | 0.8908 | 0.8908 | 0.8908 | 0.8883 | **0.8950** | 0.8908 | **0.9200** |
| F-measure | 0.9079 | **0.9082** | 0.9052 | 0.8998 | 0.9062 | 0.9055 | 0.9067 | 0.9072 | 0.9079 | 0.9063 | **0.9213** |

| | $K_p$ | | | |
|---|---|---|---|---|
| | **0 ~ 1 + and 6- ~ 9 [KN2022]** | | **2- ~ 5 + and 6- ~ 9 [This study]** | |
| | **MLP** | **PL** | **MLP** | **PL** |
| Accuracy | 0.9904 | 0.9875 | 0.9213 | 0.9100 |
| Precision | 0.9925 | 0.9867 | 0.9232 | 0.9276 |
| Recall | 0.9883 | 0.9883 | 0.9200 | 0.8900 |
| F-measure | 0.9904 | 0.9875 | 0.9213 | 0.9082 |

Summary of values of accuracy, precision, recall and F-measure (four measures) computed while varying "*r*" from 1 to 10 is shown in (a). Bold letters indicate their maxima. Four measure values in MLP (multilayer perceptron, which means normal neural network) and PL cases during extremely low- and high-geomagnetic activities as examined in KN2022, and those under moderate- and extremely high-geomagnetic conditions which were discussed in this study are listed in (b)

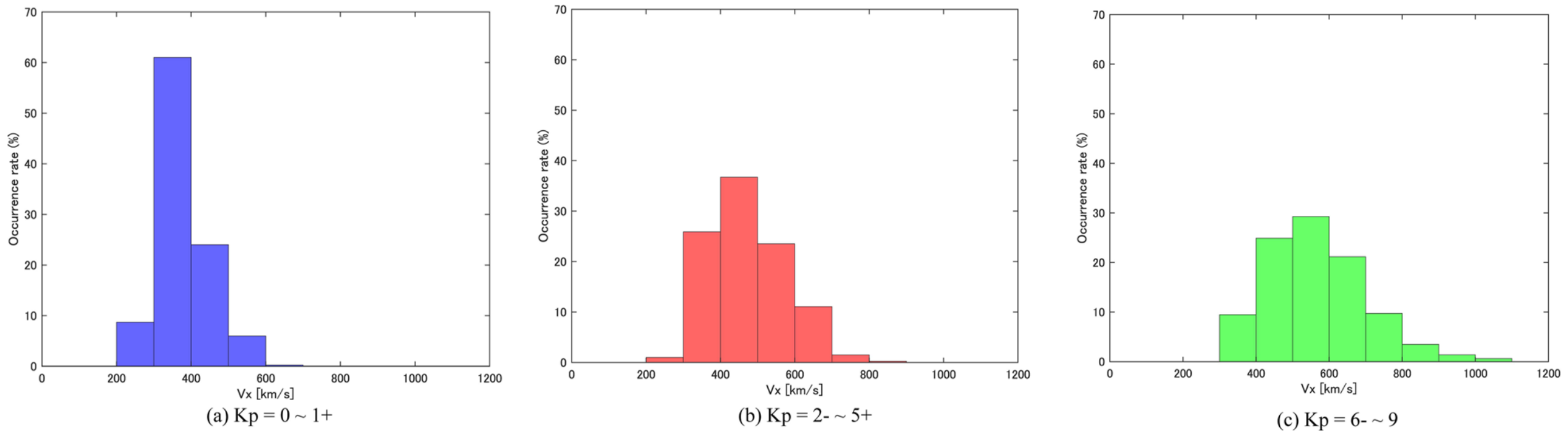

**Fig. 3** Occurrence rates of the solar wind velocity in the negative and positive target databases of PL used in Kitajima and Nowada et al. (2022, referred to as KN2022) and this study are shown. From panels a to c, histograms of the solar wind occurrence distributions under extremely low ($K_p$ = 0 to 1 +)-, moderate ($K_p$ = 2- to 5 +)-, and extremely high ($K_p$ = 6- to 9)-geomagnetic conditions are shown

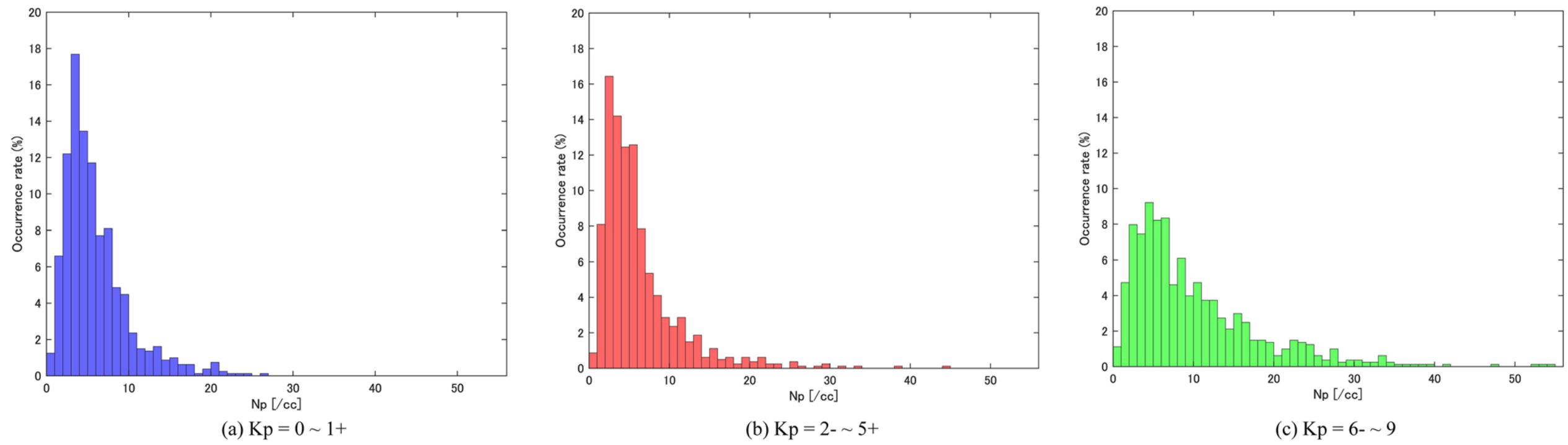

**Fig. 4** Occurrence rates of the solar wind plasma density in the negative and positive target databases of PL used in KN2022 and this study are shown. Formats of the histograms shown from panels **a** to **c** are the same as those in Fig. 3

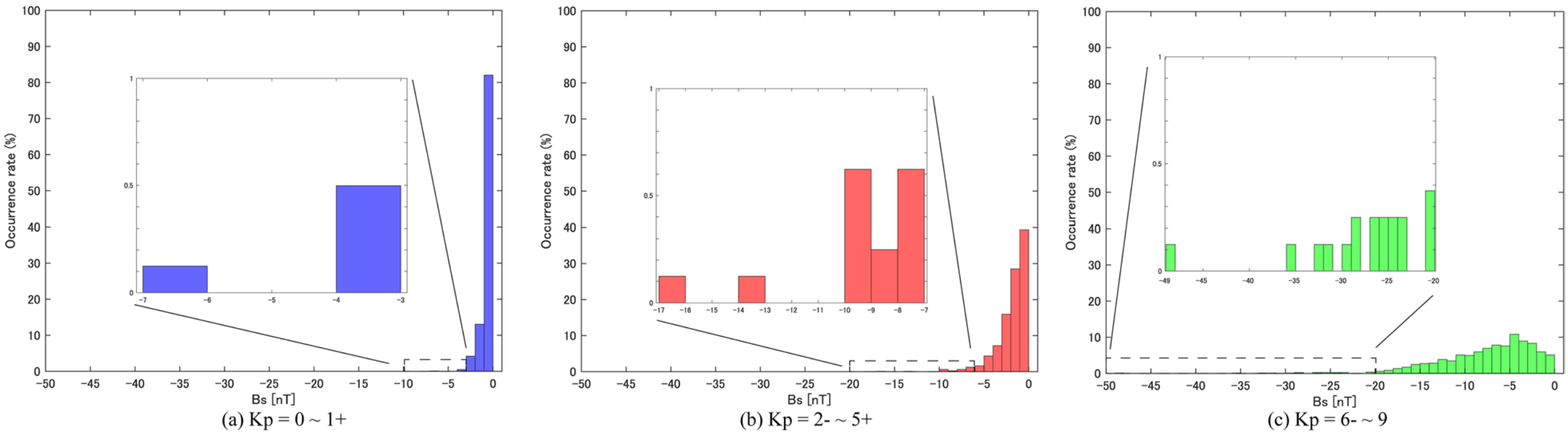

**Fig. 5** Occurrence rates of the IMF-$B_S$ (southward $B_Z$) component in the negative and positive target databases of PL used in KN2022 and this study are shown. The magnified IMF-$B_S$ histograms to clearly show all IMF-$B_S$ occurrence rates within each $K_p$ group are also displayed, although formats of the whole histograms shown from panels a to c are the same as those in Figs. 3 and 4

level. In both cases of moderate and extremely high $K_p$, the velocity peaks can be seen in the ranges of 400 km/s–500 km and 500 km/s–600 km/s, respectively. This proportional relation between solar wind velocity ranges and geomagnetic activity levels have already been revealed, based on statistical studies using large-scale databases (Newell et al. 2008) and machine learning (PL) techniques (KN2022). In the plasma density distributions shown in Fig. 4, the main peak ranges for all $K_p$ levels exhibit little variations, remaining within 10.0/cc. When the plasma density, however, exceeded 10.0/cc, at extremely low $K_p$, its occurrence rate abruptly decreased and became negligible before reaching 30.0/cc. In contrast, at moderate and extremely high $K_p$, the plasma density occurrence rate exhibited more gradual decreases but remained measurable even over 30.0/cc. In the case of extremely high $K_p$, geomagnetic condition should be supported strongly by fast solar wind plasma velocity (KN2022) while moderate-geomagnetic activity has solar wind range slower than that in the extremely high-$K_p$ case. The associated plasma density distribution is broader than that in extremely low-geomagnetic activity, suggesting that both solar wind velocity and high plasma density may contribute to the level of moderate-geomagnetic activity. This has already been suggested by Newell et al. (2008) based on an equation showing the relationship between $K_p$ and solar wind parameters. The distributions of the IMF-$B_S$ values in each $K_p$ were plotted in Fig. 5, showing that the larger the $K_p$ ranges become, the intensity of the IMF-$B_S$ component decreased, because the IMF-$B_S$ peaks and its distributions move to smaller range. These IMF-$B_S$ distributions reflect good fundamental magnetospheric energy cycle (convection), that is, the energy originating from solar wind transfers to the magnetosphere via dayside magnetic reconnection and finally contributes to enhancement of geomagnetic activity, as proposed by Dungey (1961).

It remains unclear, however, how much the plasma density contributes to geomagnetic activity levels. To investigate this, we performed an analysis based on PL and investigated the significance of the plasma density between moderate and extremely high $K_p$ levels.

### 3.3 Determination of influential solar wind parameters for moderate and extremely high $K_p$

Figure 6 shows the results of PL in KN2022 (panel a) and in this study (panel b) for the input parameters at $r=5$ (KN2022) and $r=2$ (this study). In the extremely low- and high-$K_p$ cases, PL extracted the solar wind velocity as the highest potentiality parameter (~ 1.0000), indicating that the solar wind velocity can be the most significant parameter to govern extremely low- and high-geomagnetic conditions under southward IMF conditions. The second highest potentiality was assigned to the solar wind plasma density (0.1208); however, we did not examine and discuss how the second highest potentiality parameter makes an impact on geomagnetic activity.

The solar wind velocity has the highest potentiality (1.0000), but the second highest potentiality value of the plasma density (0.4021) is 3.5 times higher than that of KN2022. This result suggests that the plasma number density can be a more influential solar wind parameter for extremely high- and moderate-geomagnetic activity than for the extremely low and high $K_p$ levels. The potentiality values of all IMF components (IMF-$B_X$, IMF-$B_Y$, and IMF-$B_S$) increased, compared with those in KN2022. The potentiality values of solar wind plasma (velocity and number density) are, however, much higher than those of the three components of IMF, indicating that intensities in each IMF component, particularly IMF-$B_S$ (southward $B_Z$) component, do not always play dominant roles in causing significant geomagnetic disturbances and governing their scale.

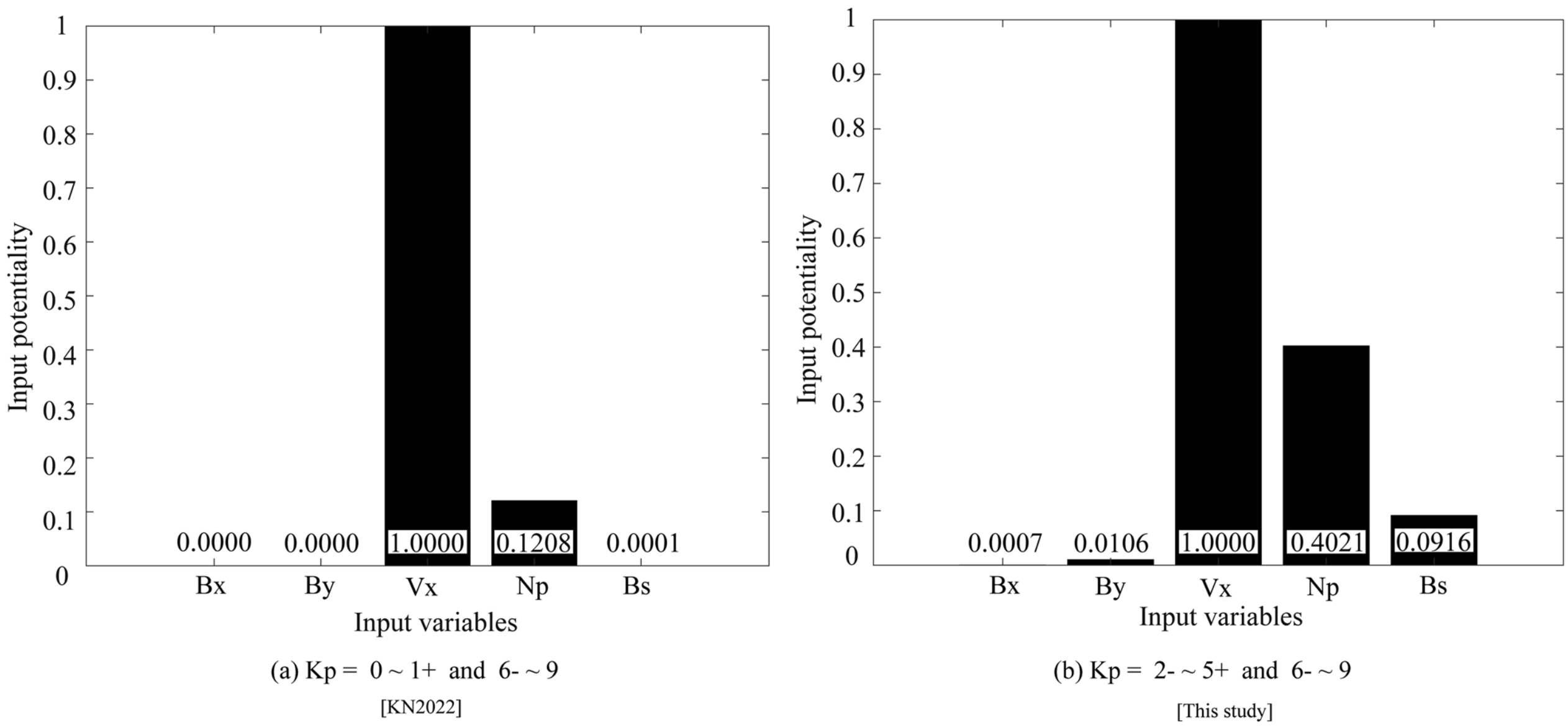

**Fig. 6** Results of PL which were obtained from KN2022 (panel **a**) and this study (panel **b**) are shown. Five OMNI solar wind parameters (IMF-$B_x$, IMF-$B_y$, $V_x$, $N_p$, and $B_s$) were chosen as PL input data. The horizontal and vertical axes represent the input potentiality and five solar wind parameters, respectively. The potentialities are shown in each histogram bar

## 4 Conclusions and discussion

In this paper, we discussed the relationship between the solar wind conditions and the moderate- and extremely high-geomagnetic disturbances. A new NN (PL) could extract the most significant solar wind parameter for the moderate and extremely high $K_p$ levels. Our PL algorithm was succeeded in revealing the dependence of solar wind conditions on extremely low- and high-geomagnetic activities in previous KN2022. In a series of our studies (this study and KN2022), we utilized 22 years' worth of normalized OMNI solar wind data obtained under southward IMF conditions as the input parameters for the PL classifier. Using the large solar wind and $K_p$ databases, PL extracted the solar wind velocity and plasma number density as the parameters with the highest and second highest potentiality, respectively, as shown in Fig. 6. Even under moderate-geomagnetic and southward IMF-$B_Z$ conditions, the solar wind velocity is the most significant parameter for moderate- and extremely high-geomagnetic activity as well as extremely low- and high-$K_p$ cases. This result suggests that the IMF-$B_S$ (southward IMF-$B_Z$) component is not always dominant parameter in causing the magnetospheric activity and governing its scale.

Many previous studies (e.g., Snyder et al. 1963; Vasyliunas et al. 1982; Borovsky et al. 1998; Gholipour et al. 2004; Newell et al. 2007, 2008; Elliott et al. 2013; KN2022, and references therein) have been discussed the dependence of solar wind speed ($V_X$) on geomagnetic activity, based on in situ observations and machine learning techniques. According to the empirical equation, that is, Eqs. (3) and (4) proposed by Newell et al. (2008), which describes the relation between the solar wind parameters and $K_p$ index and was led, based on a statistical study, $K_p$ can be expressed as follows:

$$K_p = 0.05 + 2.244 \times 10^{-4}\left(\frac{d\Phi_{MP}}{dt}\right) + 2.844 \times 10^{-6} N_P^{\frac{1}{2}} V_{SW}^2, \tag{3}$$

$$\frac{d\Phi_{MP}}{dt} = V_{SW}^{\frac{4}{3}} B_T^{\frac{2}{3}} \sin^{\frac{8}{3}}\left(\frac{\theta_{CLOCK}}{2}\right), \tag{4}$$

where $N_P$, $V_{SW}$, $B_T$, and $\theta_{CLOCK}$ indicate the solar wind plasma density, solar wind velocity, IMF intensity, and clock angle, defined by arctan (IMF-$B_Y$/IMF-$B_Z$), respectively. $K_p$ can be a function of the two solar wind velocity terms with square and 4/3 power in the solar wind convection electric field, represented by the solar wind–magnetosphere coupling function, as described in Eq. (4) (Newell et al. 2007). Therefore, the $K_p$ index is strongly correlated with the solar wind velocity.

Our PL main result shows a good agreement with this Newell's empirical relationship between $K_p$ and solar wind velocity (Fig. 6), although this result was the same as that in KN2022. In this case, however, the value of potentiality of the plasma density was higher than that in case of KN2022, indicating that at the moderate- and extremely high-$K_p$ cases, the solar wind plasma density is more significant parameter that the case at extremely low- and high-geomagnetic activity. We did not consider

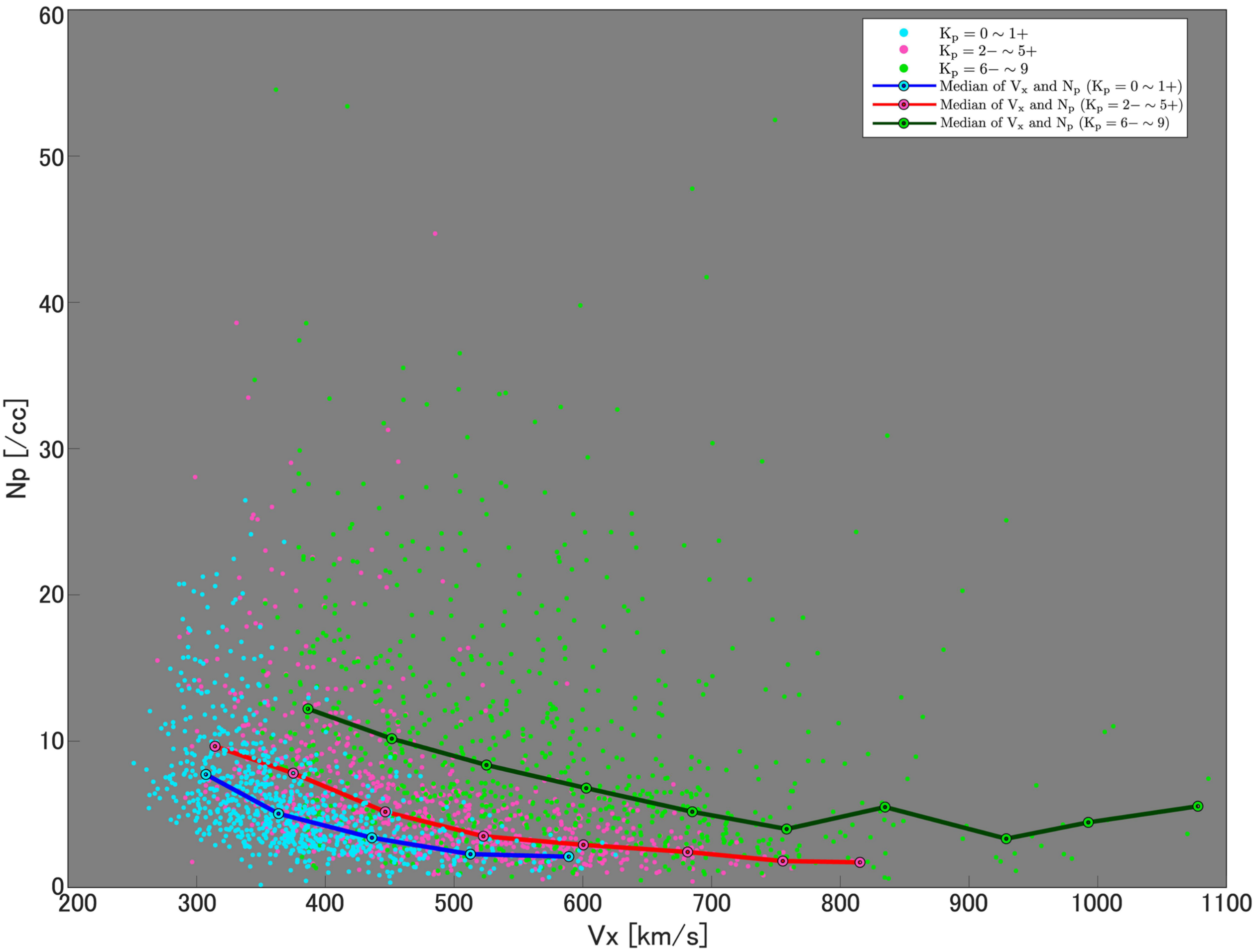

**Fig. 7** Scatter plots of the relations between solar wind velocity (horizontal axis) and solar wind plasma number density (vertical axis) in cases of extremely low (blue dots)-, moderate (magenta dots)-, and extreme high (green dots)-geomagnetic activities are shown. The median values (large two-colored circles) and their curves are also superimposed on them

to add the solar wind–magnetosphere coupling function as the input parameter of PL, because this term includes solar wind velocity and IMF intensity, which consists of the IMF-$B_X$ and -$B_Y$. These three solar wind parameters had already been selected as the PL input parameters. Therefore, we can estimate and discuss whether or not the solar wind coupling function can be significant, if taking a look at the potentiality values of these three parameters. Equation (4) shows that the coupling function mainly consists of the terms of solar wind velocity and IMF intensity. When we consider which parameter more effectively depends on coupling function, these two parameters can be considered as proxies of the solar wind dynamic pressure (kinetic energy) and magnetic pressure (magnetic energy). In general, the solar wind dynamic pressure is much more dominant than the magnetic pressure in solar wind. Furthermore, since the plasma number density is also a function of the solar wind dynamic pressure and resultant parameter with the second highest potentiality, the IMF intensity cannot become a significant parameter for (moderate) geomagnetic activity.

Figure 7 shows a summary scatter plot of the relation between the solar wind velocity (horizontal axis) and plasma density (vertical axis) in each $K_p$ level using the negative and positive target data of PL in both KN2022 and this study. The median large two-colored circles and their curves are also drawn in each $K_p$ range. All median values were calculated, based on $V_X$ from 250 to 1,130 km/s and the corresponding $N_P$ values, within a bin of 80 km/s. The $N_P$-$V_X$ profiles between extremely low (blue dots), moderate (magenta dots), and high (green dots) are different in each $K_p$ level, although the three distributions are overlapped in low-velocity and small density range. The solar wind velocity ranged from 400–1000 km/s in extremely high $K_p$, while its ranges were 300–800 km/s, and 250–600 km/s at extremely low- and moderate-geomagnetic activity. Interestingly, the larger geomagnetic disturbances become, the associated density distribution ranges are also higher. These results show that under large (small) geomagnetic activity, solar wind is fast (slow) and plasma density is also high (small) during southward IMF interval.

This study and its comparison with the result in KN2022 clarified that contribution of the plasma density to the $K_p$ index becomes more effective as the geomagnetic activity increases, being supported by the result of our Fig. 6. Particularly, we found significant differences in the $N_P$-$V_X$ relation among three $K_p$ categories over the solar wind velocity range of 300 km/s to 600 km/s (Fig. 7). Furthermore, the differences in median values between moderate and extremely high $K_p$ is larger than those between extremely low and moderate $K_p$, indicating that from the moderate-geomagnetic activity, the plasma density begins to be an influential parameter for geomagnetic activity, although the solar wind velocity is the most significant parameter for all geomagnetic activity levels, irrespectively of the $K_p$ values. These detailed results cannot be obtained from the PL investigation alone, but can be revealed by incidental statistical examinations.

Our results obtained suggest that even under southward IMF conditions, the geomagnetic disturbance level can depend on the solar wind plasma conditions with fast (slow) solar wind velocity and low (high) plasma density. Since both this study and KN2022 examined solar wind databases spanning approximately two solar cycles, it is likely that different solar wind structures during various phases may influence the intriguing PL and statistical results. If such an impact exists, our findings suggest that the relationship between $K_p$ and solar wind conditions could exhibit significant solar cycle dependence. Meanwhile, Grandin et al. (2019) reported an encounter of peculiar solar wind structures with large (small) solar wind velocity and low (high) plasma density with the Earth and the associated geoeffectiveness. The corresponding geomagnetic conditions were much (not so much) disturbed under the low (high) plasma density when solar wind velocity was fast (slow). Investigations on the relationship between the solar wind velocity and the plasma number density as our present work and KN2022 performed would also be important in forecasting the geomagnetic activity during encounters of a peculiar solar wind structures accompanied by southward IMF.

We showed that PL extracts the solar wind velocity as the most significant solar wind parameter from the solar wind database when the $K_p$ index represented moderate- and extremely high-geomagnetic activity during southward IMF intervals. Compared our results with those derived under extremely low- and high-geomagnetic activities (KN2022), however, the plasma number density has the second highest potentiality, ~ 3.5 times higher than the potentiality of the solar wind density in the KN2022 case. Although this result is based on long-term datasets spanning 22 years of solar wind parameters and $K_p$ index data, the period does not fully encompass two complete solar cycles. Therefore, as part of our future work, it is necessary to verify whether the potentiality of solar wind density varies when datasets covering two full solar cycles become available. The identical physical statistical analysis using negative and positive target data in PL reveled that the plasma number density plays a role in regulating geomagnetic disturbances at the extremely high- and moderate $K_p$ levels.

Both this study and KN2022 suggest that under southward IMF conditions, global geomagnetic activity with the $K_p$ index greater than 2 – may be closely linked to variations in solar wind velocity and plasma density. The results of this study will greatly contribute to understanding the general relationship between solar wind conditions and geomagnetic activity under various IMF conditions. When combined with future studies on the relationship between solar wind parameters and geomagnetic disturbances under northward IMF conditions, these findings will provide a more comprehensive understanding of solar wind–magnetosphere interactions.

**Abbreviations**

| | |
|---|---|
| PL | Potential learning |
| IMF | Interplanetary magnetic field |
| RMSE | Root-mean-square error |
| NN | (Artificial) neural network |
| MLP | Multilayer perceptron |
| GSM coordinates | Geocentric solar magnetospheric coordinates |
| SOM | Self-organizing map |

## Supplementary Information

The online version contains supplementary material available at https://doi.org/10.1186/s40645-025-00749-9.

Additional file 1.

**Acknowledgements**

We would like to thank Editage (www.editage.com) for English language editing in initial stage of the manuscript.

**Author contributions**

Motoharu Nowada conceived the project and wrote/edited the manuscript. Ryozo Kitajima performed all data analyses, created all the figures, and tuned the PL codes. Ryotaro Kamimura developed the main engine of the PL program. All authors critically reviewed and revised the manuscript and approved the final version for submission.

**Funding**

M.N. was supported by a grant of National High-end Foreign Experts Introduction Program (H20240313).

**Availability of data and materials**

Solar wind OMNI data were obtained from the Coordinated Data Analysis Web (https://cdaweb.sci.gsfc.nasa.gov/index.html) provided by GSFC/NASA. $K_p$

index data were provided by the World Data Center for Geomagnetism, Kyoto (http://swdcdb.kugi.kyoto-u.ac.jp/).

## Declarations

**Ethics approval and consent to participate**
Not applicable.

**Consent for publication**
Not applicable.

**Competing interests**
The authors declare that they have no competing interest.

**Author details**
[1]Department of Engineering, Tokyo Polytechnic University, 5-45-1 Iiyama-Minami, Atsugi, Kanagawa 243-0297, Japan. [2]Shandong Provincial Key Laboratory of Optical Astronomy and Solar-Terrestrial Environment, Institute of Space Sciences, Shandong University, 180 Wen-Hua West Road, Weihai 264209, Shandong, People's Republic of China. [3]Tokai University, 4-1-1 Kitakaname, Hiratsuka, Kanagawa 259-1292, Japan.

# Progress in Earth and Planetary Science
# Supplementary Information (Additional File)

## Dependence of the solar wind plasma density on moderate- and extremely high-geomagnetic activity elucidated by potential learning

Ryozo Kitajima[1‡], Motoharu Nowada[2‡*], and Ryotaro Kamimura[3]

1. Department of Engineering, Tokyo Polytechnic University, 5-45-1 Iiyama-minami, Atsugi, Kanagawa, 243-0297, Japan.

2. Shandong Provincial Key Laboratory of Optical Astronomy and Solar-Terrestrial Environment, Institute of Space Sciences, Shandong University, Weihai, Shandong, 264209, People's Republic of China.

3. Tokai University, 4-1-1 Kitakaname, Hiratsuka, Kanagawa, 259-1292, Japan.

*Correspondence to*: Ryozo Kitajima (r.kitajima@eng.t-kougei.ac.jp);
Motoharu Nowada (moto.nowada@sdu.edu.cn)

‡Equally contributed to the paper

**Contents of this file**

Table S1: Definitions of the four measures (accuracy, precision, recall, and F-measure) and formulae to compute their values are summarized.

| | | Model output | |
|---|---|---|---|
| | | Positive | Negative |
| Actual | Positive | TP | FN |
| | Negative | FP | TN |

TP: True Positive, TN: True Negative
FP: False Positive, FN: False Negative

TP: Instances correctly predicted as positive.
TN: Instances correctly predicted as negative.
FP: Instances incorrectly predicted as positive.
FN: Instances incorrectly predicted as negative.

Accuracy: a metric used to evaluate the performance of a classification model and measures the proportion of correct predictions made by the model out of the total number of predictions.

Precision: how many of the predicted positive instances are actually positive.

Recall: how many of the actual positive instances are correctly predicted.

F-measure: a performance metric used to evaluate the effectiveness of a classification model, which combines Precision and Recall into a single metric by calculating their harmonic mean.

$$\mathrm{Accuracy} = \frac{\mathrm{TP} + \mathrm{TN}}{\mathrm{TP} + \mathrm{FP} + \mathrm{FN} + \mathrm{TN}} \qquad \mathrm{Precision} = \frac{\mathrm{TP}}{\mathrm{TP} + \mathrm{FP}}$$

$$\mathrm{Recall} = \frac{\mathrm{TP}}{\mathrm{TP} + \mathrm{FN}} \qquad \mathrm{F-measure} = \frac{2 \times \mathrm{Precision} \times \mathrm{Recall}}{\mathrm{Precision} + \mathrm{Recall}}$$